\documentclass[letterpaper,twocolumn,showpacs,prl,superscriptaddress]{revtex4} 
\usepackage{graphicx,color}
\usepackage{amssymb}
\usepackage{amsmath,amsfonts,latexsym}
\usepackage{array,tabularx}
\usepackage{dcolumn}               

\definecolor{purple}{rgb}{0.4, 0, 0.4}

\begin{document}
\title{Probing Fermionic Condensates by Fast-Sweep Projection onto Feshbach Molecules} 
\author{S. Matyja\'skiewicz}
\affiliation{Department of Physics, King's College London, London WC2R 2LS,
UK} \affiliation{Department of Applied Mathematics, Warsaw University
of Life Sciences, Nowoursynowska 159, 02-776 Warsaw, Poland}
\author{M.~H. Szyma{\'n}ska}
\affiliation{Department of Physics, University of Warwick, Coventry
CV4 7AL, UK} \affiliation{Cavendish Laboratory, University of
Cambridge, Cambridge CB3 0HE, UK}
 
\author{K. G{\'o}ral}
\affiliation{Clarendon Laboratory, Department of Physics, University
  of Oxford, Parks Road, Oxford OX1 3PU, UK}  
  
\date{\today}

\begin{abstract}
Fast sweep projection onto Feshbach molecules has been widely used as a
probe of fermionic condensates. By determining the exact dynamics of a
pair of atoms in time varying magnetic fields, we calculate the number of
condensed and noncondensed molecules created after fast magnetic field
sweeps from the BCS to the BEC side of the resonances in $^{40}$K and
$^{6}$Li, for different sweep rates and a range of initial and final
fields. We discuss the relation between the initial fermionic condensate
fraction and the molecular condensate fraction measured after the sweep.
\end{abstract}
  
  
\maketitle 
  
Ever since the first realisations of fermionic condensation in
two-component gases of $^{40}$K \cite{Regal2004} and $^6$Li
\cite{Zwierlein2004} the question how these fragile states can be
accurately probed has been of great experimental importance. Since
most techniques rely on the time-of-flight expansion, during which the
Cooper pairs dissociate, indirect methods have been proposed. One of
the most common is the so called fast-sweep projection
\cite{Regal2004,Ketterle2008}, which consists in a rapid magnetic
field ramp from the BCS to the BEC side of the resonance. The key
point is to make the field variation fast with respect to the
many-body time scales thus preventing pair collisions from changing
the pairs' momentum distribution and re-thermalisation. With this
condition, the condensed molecules measured after the sweep originate
only from the Cooper pairs, already present at the initial time,
whereas the noncondensed molecules are converted only from the
noncondensed atoms. The expectation is that this allows the
information about the initial state to be easily unravelled from the
measurements of molecules created during the sweep.  The aim of this
Letter is to describe the fast sweep regime of fermion dynamics and to
determine whether and how the information about the initial condensate
can be extracted from the production efficiency and the distribution
of molecules after the sweep.

Despite its experimental relevance and fundamental interest,
understanding of the dynamics of a cold Fermi gas following variations
in the interaction strength is still relatively limited.
The early analyses of molecular production were based on overlapping
the initial pair state with the final molecular wave function
\cite{Perali2005}. This corresponds to an abrupt jump of the
magnetic field strength and thus cannot account for the sweep rate
dependence.
The analysis of time-dependent processes has so far
concentrated on the mean-field dynamics \cite{mean-field}, which by
definition ignores the noncondensed molecules. Thus, such methods are
inadequate for sweeps where large numbers of noncondensed
molecules were reported to be created during the ramp
\cite{Ketterle2008}.
Finally, in \cite{Altman} the dependence of the molecular production
on the sweep rate was estimated analytically without determining the
actual dynamics. This was done by overlapping the initial state with a
sweep-rate-dependent ``final state'' molecular wave function.
Clearly, more rigorous calculations, which would determine the
dynamics of the gas in time-varying fields and allow us to
capture both condensed and noncondensed molecules created during this
process, are still missing.

Making use of a systematic cumulant expansion \cite{Thorsten2002}, one
can obtain the dynamic equations for the density matrix and the pair
function of the gas. On this basis, the number of molecules beyond the
mean-field level, i.e. accounting for noncondensed pairs, can be
calculated. In this Letter we numerically solve these equations for a
limited case of fast sweeps, during which pair collisions are
insignificant. In such a case, the dynamical aspects are captured by
the two-body time evolution of a single atom pair in the spirit of the
original rationale behind the fast sweep projection technique.
We determine the two-body time evolution exactly by numerically
solving the time-dependent Schr\"{o}dinger equation for a pair of atoms.
We start from the single-channel Hamiltonian
\begin{equation}
\hat{H}=\sum_{\mathbf{k}s} \epsilon_{\mathbf{k}} 
a^{\dagger}_{\mathbf{k}s} a^{}_{\mathbf{k}s}
+\sum_{\bf{k}\bf{k'}\bf{q}} V_{{\bf{k}}{\bf{k}}'}
a^{\dagger}_{{\bf{k}}+{\bf{q}}\uparrow} 
a^{\dagger}_{-\bf{k}\downarrow} a^{}_{-{\bf{k}}'\downarrow} 
a^{}_{{\bf{k}}'+{\bf{q}}\uparrow},
\label{HMB1ch}
\vspace{-1mm}
\end{equation}
with Fermi operators $a^{\dagger}_{\mathbf{k}s}$ and
$a^{}_{\mathbf{k}s}$ (hereafter $\bf{k}$, $\bf{k}'$, and $\bf{q}$
denote wave vectors in three dimensions).  In the following we use a
finite-range potential of the form
$V_{{\bf{k}}{\bf{k}}'}=V_0(B)\chi({\bf k})\chi({\bf k}')$. Here
$\chi({\bf k})=\exp[-({\bf k}\sigma_{\mathrm{bg}})^2/2]$ and the
parameters $V_0(B)$ and $\sigma_{\mathrm{bg}}$ are chosen to recover
the magnetic field dependence of the scattering length
$a(B)=a_{\mathrm{bg}}(1-\frac{\Delta B}{B-B_0})$ and of the highest
vibrational bound state energy beyond the universal regime (details in
\cite{Goral2004,Thorsten2006,me_therm}).
The density of molecules at time $t$ \cite{Thorsten2003} is 
\vspace{-3mm}
\begin{multline}
\label{Nb}
n_m({\bf{q}},t) = 
\int d{\bf{k}}d{\bf{k'}}  \phi^{\star}_\mathrm{b}(B,\mathbf{k})
\phi_\mathrm{b}(B,\mathbf{k'})    \\
\langle a^{\dagger}_{{\bf{k}}+{\bf{q}}\uparrow} 
a^{\dagger}_{-\bf{k}\downarrow} a^{}_{-{\bf{k}}'\downarrow} 
a^{}_{{\bf{k}}'+{\bf{q}}\uparrow} \rangle_{t},
\end{multline}
where $\phi_\mathrm{b}(B,\mathbf{k})$ is the Feshbach molecule wave
function (determined by solving the stationary Schr\"{o}dinger
equation for a pair of atoms with the finite-range potential
$V_{{\bf{k}}{\bf{k}}'}$) and the time dependence of the field
is given by $B(t)=B_0+\dot{B}(t-t_0)$. 
To calculate (\ref{Nb}) with the
time-dependent Hamiltonian (\ref{HMB1ch}) we use the cumulant
expansion \cite{Thorsten2002}. In the presence of a condensate, higher
order correlation functions are still large, whereas higher order
cumulants, which include the equivalent order of interactions around
the interaction-free evolution, are small -- at least at short
times. The expansion is thus controlled by how fast the resonance is
crossed.
Having expressed (\ref{Nb}) in terms of cumulants, we can close the
hierarchy of their dynamical equations by neglecting all fifth- and
higher-order terms (the second-order expansion). 
Note, however, that since here $\langle a^{\dagger}_{\bf{k}s}
\rangle=\langle a^{}_{\bf{k}s} \rangle=0$, we need to keep only the
second-order cumulants (which here are equivalent to the correlation
functions: the density matrix $\Hat{\Gamma}({\bf
k},t)=\frac{1}{2}\sum_s\langle a^{\dagger}_{\bf{k}
s}(t)a^{}_{\bf{k}s}(t)\rangle$ and the pair function
$\Hat{\Phi}({\bf{k}},t)=\langle a^{}_{-\bf{k}
\downarrow}(t)a^{}_{\bf{k} \uparrow}(t)\rangle$) and the fourth-order
ones of the form $\langle F_1F_2F_3F_4 \rangle^c$ (where $F_i$
indicate any fermionic creation and annihilation operators).  
Next, by moving to the interaction picture, it is possible
\cite{Thorsten2002} to formally solve for all fourth-order cumulants
and to obtain close equations for $\Hat{\Gamma}({\bf k},t)$ and
$\Hat{\Phi}({\bf{k}},t)$. This results in additional non-Markovian
collision terms containing products of $\Hat{\Gamma}({\bf{k}},t)$ at
different momenta and times in addition to terms already present in
the mean-field equations \cite{mean-field}. Consequently, the
molecular density becomes
\begin{align}
\label{Nblong}
&n_m(t) = 
\left|\int d\mathbf{k}
|\langle \phi_\mathrm{b}(t) | \mathbf{k} \rangle|^2
\hat\Phi(\mathbf{k},t)\right|^2 + \\ 
& 2 \{ \int d\mathbf{k'} 
\left|T(\mathbf{k'},t,t_0)\right|^{2} \int d\mathbf{k}
\hat\Gamma(\mathbf{k},t_0) \hat\Gamma(\mathbf{k}-2\mathbf{k'},t_0) + 
\notag \\ 
& \int^t_{t_0} dt' \int d\mathbf{k'}
\left|T(\mathbf{k'},t,t')\right|^{2} \int d\mathbf{k}
\frac{\partial}{\partial t'}[\hat\Gamma(\mathbf{k},t')
\hat\Gamma(\mathbf{k}-2\mathbf{k'},t')] \}.  \notag
\end{align}
The two-body transition amplitude is defined as
$T(\mathbf{k},t,t_0) = \langle \phi_\mathrm{b}(t) |
U_\mathrm{2B}(t,t_0)|\mathbf{k} \rangle,
\label{transitionamplitude}$
where $U_\mathrm{2B}(t,t_0)$ is a two-body evolution operator. The
first term in (\ref{Nblong}) describes condensed molecules and the
remaining terms the noncondensed molecules created after time $t$. 
Second-order cumulant expansion includes two-particle collisions but
neglects higher order collision terms.  Certainly, for very slow
ramps, for which the re-thermalisation and multiple collisions are
important, a higher order expansion may be needed.  However, already
the solution of full non-Markovian second order cumulant equations is
numerically challenging and is beyond the scope of this Letter
(although a suitable parallelisation of the computation should allow
slower field variations to be studied).  Here, our intention is to
focus on the fast sweep limit. On such short time scales the evolution
of the pair function $\Hat{\Phi}$ is principally captured by the
two-body evolution operator $U_\mathrm{2B}$, i.e
$\Hat{\Phi}(t)=U_\mathrm{2B}(t,t_0)\Hat{\Phi}(t_0)$, and the third
(collision) term in equation (\ref{Nblong}) is small. In this limit
the density of condensed molecules becomes
\begin{equation}
n^c_m(t)=
\left|\int
d\mathbf{k} T(\mathbf{k},t,t_0)
\hat\Phi(\mathbf{k},t_0)\right|^2,
\label{Nbc}
\end{equation}  
where $\Hat\Phi(t_0)$ is the initial pair function. 
Density of noncondensed molecules is
\begin{equation}
\hspace{-0.02cm}n^{nc}_m=2
\hspace{-0.1cm}\int
\hspace{-0.1cm}d\mathbf{k'}
\left|T(\mathbf{k'},t,t_0)\right|^{2}\hspace{-0.1cm} \int d\mathbf{k}
\hat\Gamma(\mathbf{k},t_0) \hat\Gamma(\mathbf{k}-2\mathbf{k'},t_0),
\label{Nbnc}
\end{equation} 
where $\Hat\Gamma(t_0)$ is the initial density matrix. We determine
$U_{2B}(t,t_0)$ exactly by numerically solving $i\hbar
\frac{\partial}{\partial t} U_{2B}(t,t_0)= \hat{H}_{2B}(t)
U_{2B}(t,t_0)$, where $\hat{H}_{2B}$ is equal to (\ref{HMB1ch}) for a
pair of atoms.  We are then in a position to evaluate $n^c_m$ and
$n^{nc}_m$ from (\ref{Nbc}) and (\ref{Nbnc}), respectively. Since our
main focus is on establishing the relation between the final state
after a fast sweep and the initial state, we take the simple
mean-field thermodynamic initial conditions.
 
\begin{figure}[b]
\begin{center}
\includegraphics[width=1.0\linewidth,angle=0,clip]{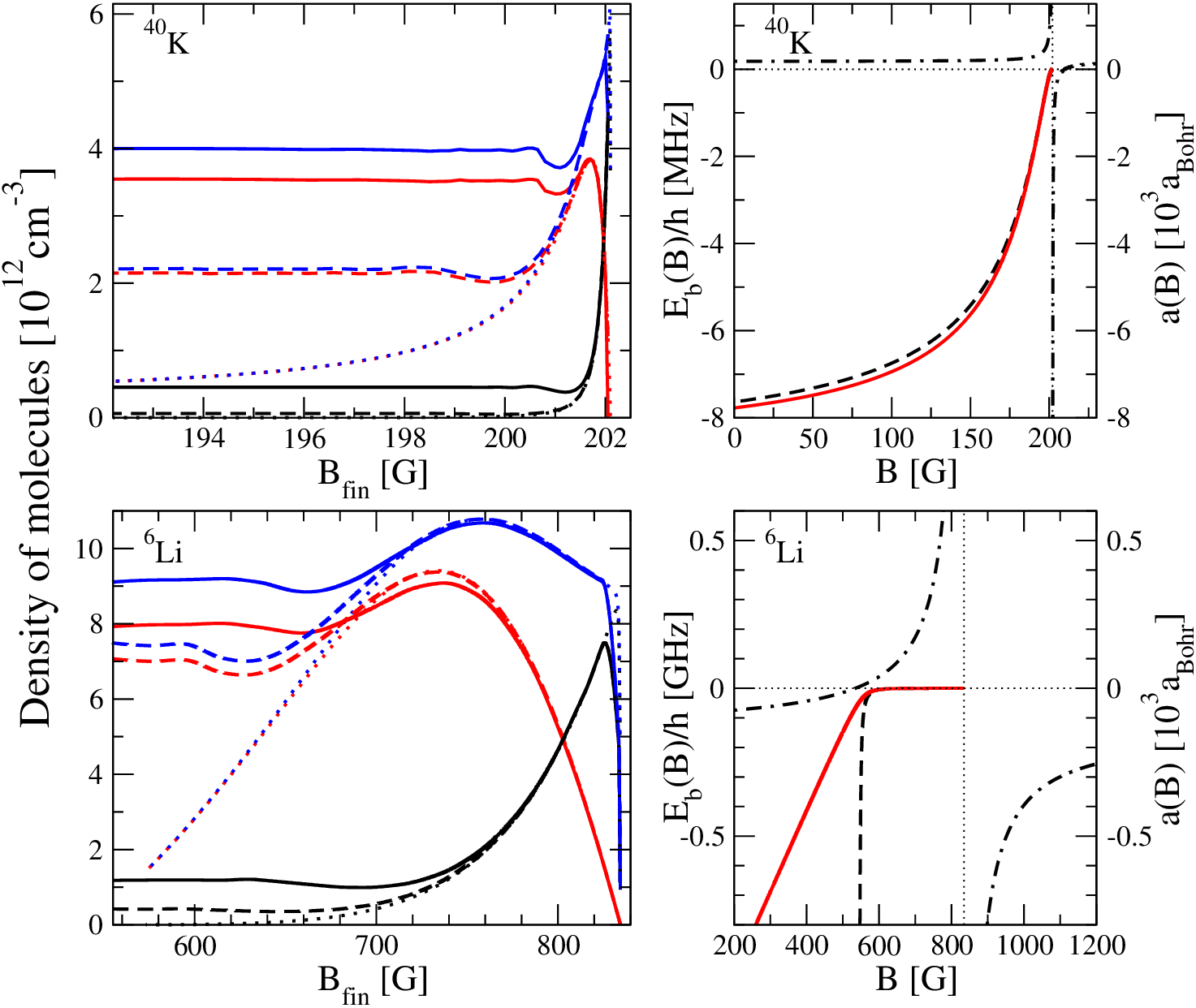}
\end{center}
\vspace{-5mm} 
\caption{(color online). Right: The scattering length (dashed-doted
        lines) and the bound state energy, $E_\mathrm{b}(B)$, of the
        $^{40}$K (top) and the $^6$Li (bottom) Fesh\-bach molecule
        $\phi_\mathrm{b}(B)$ in the vicinity of the 202 G ($^{40}$K)
        and the 834 G ($^6$Li) resonances from two- (solid lines) and
        single-channel approaches (dashed lines).  Left: Total (top
        curves, blue), condensed (middle curves, red) and noncondensed
        (bottom curves, black) molecular densities versus the final
        field, $B_{\rm{fin}}$, for the initial field
        $B_{\rm{ini}}-B_0=0.12$ G for $^{40}$K and
        $B_{\rm{ini}}-B_0=4.62$ G for $^6$Li and slew rates
        ($1/\dot{B}$): 10 $\mu\rm{s/G}$ ($^{40}$K), 0.01 $\mu\rm{s/G}$
        ($^6$Li) (solid lines); 1 $\mu\rm{s/G}$ ($^{40}$K), 0.003
        $\mu\rm{s/G}$ ($^6$Li) (dashed lines); and abrupt jump (dotted
        lines). The initial atomic densities are 1.5 ($^{40}$K) and
        2.9 ($^6$Li) $10^{13} \rm{cm}^{-3}$.  For $^{40}$K the
        parameters are from \cite{Regal2004} and for $^6$Li are
        motivated by \cite{Ketterle2008,Salomon2005}.  }
\label{fig:BfinK}
\end{figure}
We first analyse the dependence of $n^{c}_m$, $n^{nc}_m$ and
$n_m=n^{c}_m + n^{nc}_m$ on the final field (Fig. \ref{fig:BfinK}).
For finite-rate sweeps (unlike for the abrupt jump) the number of
produced molecules depends on the final field only if it is in the
region close to the resonance -- it saturates further away from it;
and the faster the sweep the further from the resonance this occurs.
This supports the picture used in \cite{Altman} that sufficiently far
from the resonance $E_b$ is so large and interactions so small that
the molecular state adiabatically follows the ramp (no molecules are
created or dissociated during this part of the dynamics)
\cite{Ketterle2008}.
For final fields near the resonance the molecular production is
independent of the sweep rate and close to the one after 
jump. There is, however, an intermediate region of fields for which
the molecular production is not yet saturated, but it depends on the
rate.  The presence of this region and also the exact position of the
crossover to adiabatic dynamics is expected to be the main source of
quantitative discrepancy between the dynamics presented here and
analytical estimates from \cite{Altman} for condensed pairs
(additional differences arise for noncondensed ones -- these are
discussed later).
Note, that there is a qualitative difference between the 202 G
resonance in $^{40}$K and the 834 G resonance in $^{6}$Li
\cite{me_therm,Duan} (see Fig \ref{fig:BfinK}). For $^{6}$Li and slew
rate $1/\dot{B} = 0.003 \mu$s/G \footnote{Note that the relevant
dimensionless slew rate which characterises the ramp speed is roughly
$\frac{\hbar n \Delta B a_{bg}}{m} \frac{1}{\dot{B}}$ \cite{Greene}
and thus $0.003 \mu$s/G for $^{6}$Li corresponds to around $12 \mu$s/G
for $^{40}$K.} used in \cite{Salomon2005} molecular production
saturates around 250 G below $B_0$, where the single and two-channel
predictions already differ (there is around 15$\%$ difference in $E_b$
between the two). This suggests that if we were to study sweeps even
faster than those in \cite{Salomon2005} an extension to two-channels
would be necessary. In contrast, for $^{40}$K $E_b$ from the single-
and two-channel approaches coincide in the relevant region $B-B_0>
-10$ G and are close for any field. Thus, the single channel theory is
sufficient for analysis of molecular production in $^{40}$K for any
sweep rate.

\begin{figure}[h]
\begin{center}
\includegraphics[width=1.0\linewidth,angle=0]{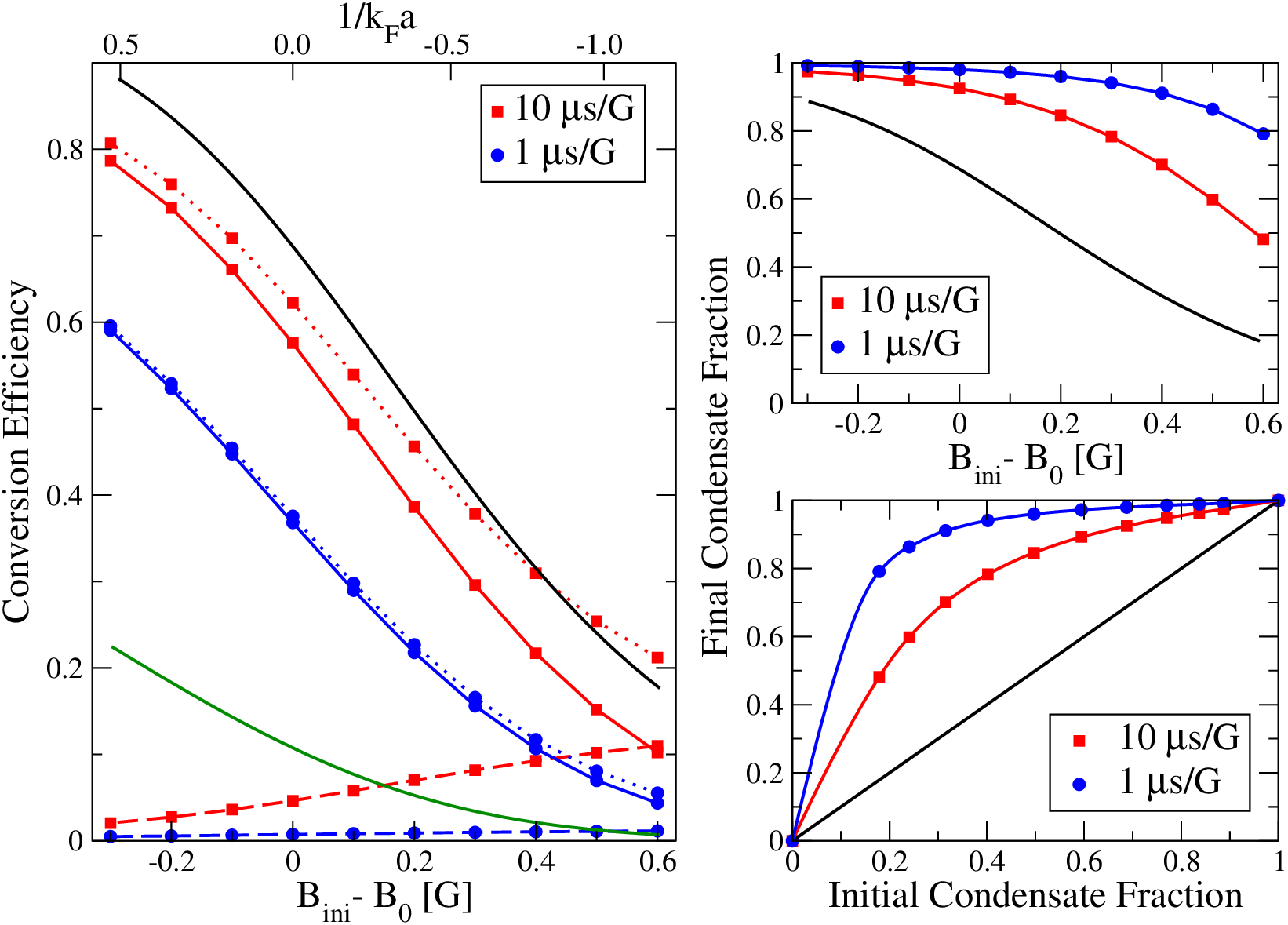}
\end{center}
\vspace{-6mm} 
\caption{(color online). Left: Molecular condensate, 2$n^c_m/n$,
   (solid lines), non-condensate, 2$n^{nc}_m/n$, (dashed lines) and
   total, 2$n_m/n$, (doted lines) conversion efficiencies versus the
   initial field, $B_{\rm{ini}}$, for the final field
   $B_{\rm{fin}}-B_0=\rm{-10}$ G and two slew rates; and 2$n^c_m/n$
   for an abrupt jump, which produces virtually no noncondensed
   molecules (bottom solid curve, green). Top solid curve (black)
   shows the initial (at $B_{\rm{ini}}$) condensate fraction,
   $2n_c/n$.  Right: Final molecular condensate fraction, $
   n^c_m/n_m$, for the same parameters versus $B_{\rm{ini}}$ (top),
   and versus the initial $2n_c/n$ (bottom).  Bottom solid curves
   (black) show the initial $2n_c/n$ . Calculations are for $^{40}$K
   202 G resonance and atomic density of 1.5 $10^{13} \rm{cm}^{-3}$.}
\label{fig:Bini}
\end{figure}
The dependence of $n^c_m$ on the initial field (Fig. \ref{fig:Bini})
is related to the number of Cooper pairs, $n_c(B)=\int d{\bf k}
|\Hat{\Phi}({\bf k},B)|^2$, at that field, but it is different for
different sweep rates.
As expected, the number of condensed molecules, $n^c_m$, created
during a linear sweep is always smaller than the number of initial
Cooper pairs but larger than what one would obtain after an abrupt
jump. The number of noncondensed molecules, $n^{nc}_m$, increases as
the initial field is shifted deeper to the BCS side.
This can be understood by noting that, in the fast sweep regime, the
noncondensed pairs are created only from atoms which were initially
out of the condensate and the number of these atoms increases towards
the BCS side.

\begin{figure}[h]
\begin{center}
\includegraphics[width=1.0\linewidth,angle=0]{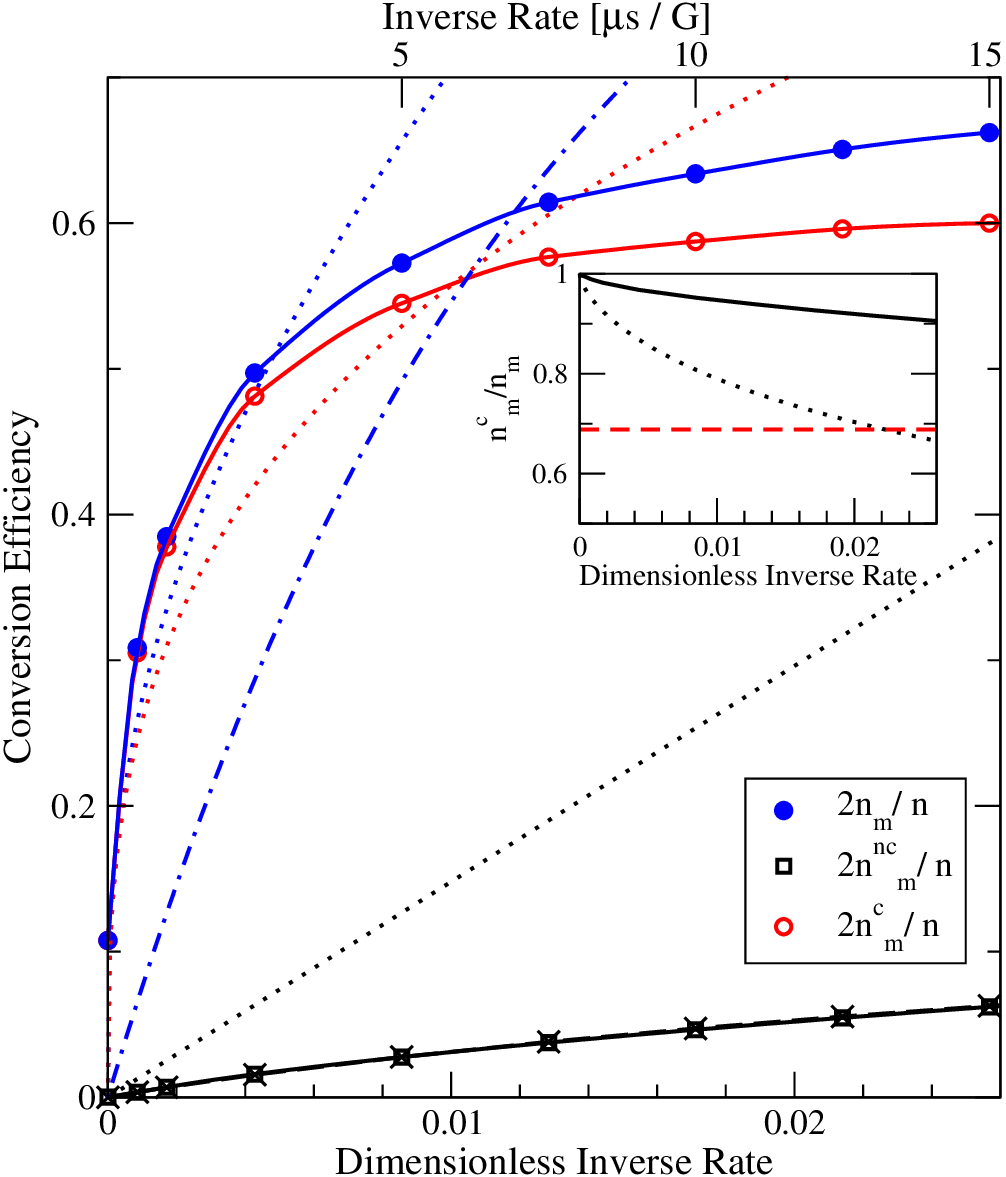}
\end{center}
\vspace{-5mm} 
\caption{(color online). Conversion efficiency for condensed,
   2$n^c_m/n$, noncondensed, 2$n^{nc}_m/n$, and total, 2$n_m/n$,
   molecules versus the dimensionless inverse sweep rate $\frac{\hbar
   n \Delta B a_{bg}}{m} \frac{1}{\dot{B}}$ for the $^{40}$K 202 G
   resonance and atomic density of 1.5 $10^{13} \rm{cm}^{-3}$ for
   $B_{\rm{ini}}=B_0$ and $B_{\rm{fin}}-B_0$ = -10 G. For comparison,
   formulas from \cite{Altman} (dotted lines: total, condensed and
   noncondensed from top to bottom); and for 2$n_m/n$ the Landau-Zener
   prediction from \cite{Ketterle2008} (dashed-dotted line) are
   shown. x symbols indicate the asymptotic approximation as discussed
   in the text. Inset: Final molecular condensate fraction,
   $n^{c}_m/n_m$, (solid line) for parameters as in the main Fig. For
   comparison, $n^{c}_m/n_m$ from \cite{Altman} (dotted line) and the
   initial condensate fraction, 2$n^c/n$, (flat dashed line) are
   plotted.}
\label{fig:fields_vs_rate}
\vspace{-5mm} 
\end{figure}

Finally, the dependence on the sweep rate is shown in Fig.
\ref{fig:fields_vs_rate} and compared with \cite{Altman} and, for
$n_m$, with the Landau-Zener (LZ) theory from \cite{Ketterle2008}.
Since Cooper pairs convert to molecules more efficiently than free
atoms (as pointed out in \cite{Altman}), LZ, which does not
distinguish between the two, is unlikely to give an accurate account
of molecular production from a partially condensed Fermi gas.  Indeed,
LZ results differ substantially from the results of our calculations
(see Fig. \ref{fig:fields_vs_rate}). The differences with
\cite{Altman} are pronounced especially for the production efficiency
of noncondensed molecules, which in \cite{Altman} has a linear
dependence on the slew rate with a coefficient dependent on the
density but independent of the initial field. Here, it is instructive
to compare our dynamics with an approach based on the asymptotic
dissociation spectrum given by (B12) in \cite{Hanna2006}. It can be
shown \cite{Hanna2006} that in the asymptotic limit of $t_0
\rightarrow -\infty$ and $t \rightarrow +\infty$ the transition
probability is given by $|T(\mathrm{k})|^2=\frac{|a_\mathrm{bg} \Delta
B|}{\pi \hbar^2 m |\dot{B}|} \exp\left(-\frac{4}{3}\frac{a_\mathrm{bg}
\Delta B}{\hbar^2 m \dot{B}} \mathrm{k}^3\right)$.  Using this form
instead of the exact one in (\ref{Nbnc}) gives an approximate number
of noncondensed molecules, good for $B_{\rm{ini}}$ deeply in the BCS
regime.  An interesting case is that of $B_{\rm{ini}}=B_0$ for which
$n^{nc}_m$ obtained using the exact transition probability,
$\left|T(\mathbf{k},t,t_0)\right|^{2}$, turns out to be a half of
$n^{nc}_m$ obtained using the asymptotic spectrum. On the basis of
numerical evidence we can thus provide a simple formula for the number
of noncondensed molecules for $B_{\rm{ini}}=B_0$ as a function of the
slew rate 
\begin{multline}
n^{nc}_m \approx
\frac{|a_\mathrm{bg} \Delta B|}{2\pi \hbar^2 m |\dot{B}|}\int
d\mathbf{k'}\exp\left(\hspace{-0.05cm}-\frac{4}{3}\frac{a_\mathrm{bg}
\Delta B}{\hbar^2 m \dot{B}}\mathrm{k'}^3\right)  \\
\int  d\mathbf{k} \hat\Gamma(\mathbf{k},t_0)
\hat\Gamma(\mathbf{k}-2\mathbf{k'},t_0)
\end{multline}
(x symbols in Fig. \ref{fig:fields_vs_rate}), whose agreement with the
exact one is very good indeed. 
The dependence on $1/|\dot{B}|$ is a product of a linear term,
dominant for very fast sweeps, and an exponential correction.  The
coefficient of this dependence is determined by the overlap of the
transition probability and the initial density matrix at different
momenta, and thus varies with the initial field (see
Fig. \ref{fig:Bini}).

As shown in Fig. \ref{fig:Bini} and \ref{fig:fields_vs_rate} (Inset),
the molecular condensate fraction, $n^c_m/n_m$, is always much larger
than the initial one, $2n_c/n$, but also it decreases to some extent
with the slew rate. However, for slower sweeps the many-body effects
become more important, which would again lead to an increase of the
condensate fraction by formation of new condensed molecules from
initially noncondensed atoms. This effect, together with three- and
four-body collision losses which are density dependent and thus larger
for condensed pairs occupying a smaller region in the centre of the
trap \cite{Martin}, is a possible explanation for the sweep rate
independent measurement of the condensate fraction in $^{40}$K (Fig. 5
in \cite{Regal2004}). Note, that only the fastest sweeps in this
Figure are expected to be in the fully two-body regime. Due to the
large interchannel coupling it has proven to be much harder to realise
the fast sweep limit for $^{6}$Li. For the fastest sweeps reported
\cite{Salomon2005} of 0.003 $\mu\rm{s/G}$ (dashed lines in the left
bottom panel of Fig. \ref{fig:BfinK}), atomic density $n$=2.9 $10^{13}
\rm{cm}^{-3}$ and initial field around 4G above the resonance, our
method predicts around 0.49 for $2n^{c}_m/n$ and about 0.03 for
$2n^{nc}_m/n$.
   
To conclude, we have shown that even in the limit of fast sweeps,
where the dynamics of the gas is predominantly governed by the
two-body evolution, the number of condensed molecules as well as the
molecular condensate fraction after the sweep depend on both the sweep
rate and the initial state.  This makes it difficult to unravel the
initial condensate fraction (and the initial number of Cooper pairs)
from the measurement of the final molecular condensate without
detailed dynamical calculations. However, although in the fast sweep
limit the measured molecular condensate fraction always overestimates
the initial condensate fraction, the presence of the molecular
condensate implies the existence of the fermionic condensate before
the sweep.

We are grateful to T. K{\"o}hler, K. Burnett, T. M. Hanna,
A. F. G{\'o}ral and M. W. Zwierlein for stimulating discussions. This
research has been supported by Marie-Curie fellowship (S. M.), and
EPSRC (M. H. S.).
  
\bibliography{biblio}

\newcommand\textdot{\.}
\begin{thebibliography}{22}
\expandafter\ifx\csname natexlab\endcsname\relax\def\natexlab#1{#1}\fi
\expandafter\ifx\csname bibnamefont\endcsname\relax
  \def\bibnamefont#1{#1}\fi
\expandafter\ifx\csname bibfnamefont\endcsname\relax
  \def\bibfnamefont#1{#1}\fi
\expandafter\ifx\csname citenamefont\endcsname\relax
  \def\citenamefont#1{#1}\fi
\expandafter\ifx\csname url\endcsname\relax
  \def\url#1{\texttt{#1}}\fi
\expandafter\ifx\csname urlprefix\endcsname\relax\def\urlprefix{URL }\fi
\providecommand{\bibinfo}[2]{#2}
\providecommand{\eprint}[2][]{\url{#2}}

\bibitem[{\citenamefont{Regal et~al.}(2004)\citenamefont{Regal, Greiner, and
  Jin}}]{Regal2004}
\bibinfo{author}{\bibfnamefont{C.~A.} \bibnamefont{Regal}},
  \bibinfo{author}{\bibfnamefont{M.}~\bibnamefont{Greiner}}, \bibnamefont{and}
  \bibinfo{author}{\bibfnamefont{D.~S.} \bibnamefont{Jin}},
  \bibinfo{journal}{Phys. Rev. Lett.} \textbf{\bibinfo{volume}{92}},
  \bibinfo{eid}{040403} (\bibinfo{year}{2004}).

\bibitem[{\citenamefont{Zwierlein et~al.}(2004)\citenamefont{Zwierlein, Stan,
  Schunck, Raupach, Kerman, and Ketterle}}]{Zwierlein2004}
\bibinfo{author}{\bibfnamefont{M.~W.} \bibnamefont{Zwierlein \emph{et al.}}},
  \bibinfo{journal}{Phys. Rev. Lett.} \textbf{\bibinfo{volume}{92}},
  \bibinfo{eid}{120403} (\bibinfo{year}{2004}).

\bibitem[{\citenamefont{for review~see W.~Ketterle and
  Zwierlein}(2008)}]{Ketterle2008}
\bibinfo{author}{\bibnamefont{for review~see W.~Ketterle}} \bibnamefont{and}
  \bibinfo{author}{\bibfnamefont{M.~W.} \bibnamefont{Zwierlein}},
  \bibinfo{journal}{arXiv:0801.2500 and references there in}
  (\bibinfo{year}{2008}).

\bibitem[{\citenamefont{Perali et~al.}(2005)\citenamefont{Perali, Pieri, and
  Strinati}}]{Perali2005}
\bibinfo{author}{\bibfnamefont{R. B.}~\bibnamefont{Diener}},
  \bibinfo{author}{\bibfnamefont{T. L.}~\bibnamefont{Ho}},
\bibinfo{journal}{cond-mat/0404517} (\bibinfo{year}{2004}),
\bibinfo{author}{\bibfnamefont{A.}~\bibnamefont{Perali}},
  \bibinfo{author}{\bibfnamefont{P.}~\bibnamefont{Pieri}}, \bibnamefont{and}
  \bibinfo{author}{\bibfnamefont{G.~C.} \bibnamefont{Strinati}},
  \bibinfo{journal}{Phys. Rev. Lett.} \textbf{\bibinfo{volume}{95}},
  \bibinfo{eid}{010407} (\bibinfo{year}{2005}).

\bibitem[{\citenamefont{Barankov and Levitov}(2004)}]{mean-field}
\bibinfo{author}{\bibfnamefont{R.~A.} \bibnamefont{Barankov}} \bibnamefont{and}
  \bibinfo{author}{\bibfnamefont{L.~S.} \bibnamefont{Levitov}},
  \bibinfo{journal}{Phys. Rev. Lett.} \textbf{\bibinfo{volume}{93}},
  \bibinfo{pages}{130403} (\bibinfo{year}{2004});
\bibinfo{author}{\bibfnamefont{R.~A.} \bibnamefont{Barankov}},
  \bibinfo{author}{\bibfnamefont{L.~S.} \bibnamefont{Levitov}},
  \bibnamefont{and} \bibinfo{author}{\bibfnamefont{B.~Z.}
  \bibnamefont{Spivak}}, \bibinfo{journal}{Phys. Rev. Lett.}
  \textbf{\bibinfo{volume}{93}}, \bibinfo{pages}{160401}
  (\bibinfo{year}{2004});
\bibinfo{author}{\bibfnamefont{A.~V.} \bibnamefont{Andreev}},
  \bibinfo{author}{\bibfnamefont{V.}~\bibnamefont{Gurarie}}, \bibnamefont{and}
  \bibinfo{author}{\bibfnamefont{L.}~\bibnamefont{Radzihovsky}},
  \bibinfo{journal}{Phys. Rev. Lett.} \textbf{\bibinfo{volume}{93}},
  \bibinfo{pages}{130402} (\bibinfo{year}{2004});
\bibinfo{author}{\bibfnamefont{M.~H.} \bibnamefont{Szyma\'{n}ska}},
  \bibinfo{author}{\bibfnamefont{B.~D.} \bibnamefont{Simons}},
  \bibnamefont{and} \bibinfo{author}{\bibfnamefont{K.}~\bibnamefont{Burnett}},
  \bibinfo{journal}{Phys. Rev. Lett.} \textbf{\bibinfo{volume}{94}},
  \bibinfo{eid}{170402} (\bibinfo{year}{2005}{\natexlab{a}});
\bibinfo{author}{\bibfnamefont{E.~A.} \bibnamefont{Yuzbashyan}},
  \bibinfo{author}{\bibfnamefont{V.~B.} \bibnamefont{Kuznetsov}},
  \bibnamefont{and} \bibinfo{author}{\bibfnamefont{B.~L.}
  \bibnamefont{Altshuler}}, \bibinfo{journal}{Phys. Rev. B}
  \textbf{\bibinfo{volume}{72}}, \bibinfo{eid}{144524} (\bibinfo{year}{2005});
\bibinfo{author}{\bibfnamefont{E.~A.} \bibnamefont{Yuzbashyan}},
  \bibinfo{author}{\bibfnamefont{O.}~\bibnamefont{Tsyplyatyev}},
  \bibnamefont{and} \bibinfo{author}{\bibfnamefont{B.~L.}
  \bibnamefont{Altshuler}}, \bibinfo{journal}{Phys. Rev. Lett.}
  \textbf{\bibinfo{volume}{96}}, \bibinfo{eid}{097005} (\bibinfo{year}{2006});
\bibinfo{author}{\bibfnamefont{E.~A.} \bibnamefont{Yuzbashyan}}
  \bibnamefont{and} \bibinfo{author}{\bibfnamefont{M.}~\bibnamefont{Dzero}},
  \bibinfo{journal}{Phys. Rev. Lett.} \textbf{\bibinfo{volume}{96}},
  \bibinfo{eid}{230404} (\bibinfo{year}{2006}).

\bibitem[{\citenamefont{Altman and Vishwanath}(2005)}]{Altman}
\bibinfo{author}{\bibfnamefont{E.}~\bibnamefont{Altman}} \bibnamefont{and}
  \bibinfo{author}{\bibfnamefont{A.}~\bibnamefont{Vishwanath}},
  \bibinfo{journal}{Phys. Rev. Lett.} \textbf{\bibinfo{volume}{95}},
  \bibinfo{eid}{110404} (\bibinfo{year}{2005}).

\bibitem[{\citenamefont{K\"ohler and Burnett}(2002)}]{Thorsten2002}
\bibinfo{author}{\bibfnamefont{T.}~\bibnamefont{K\"ohler}} \bibnamefont{and}
  \bibinfo{author}{\bibfnamefont{K.}~\bibnamefont{Burnett}},
  \bibinfo{journal}{Phys. Rev. A} \textbf{\bibinfo{volume}{65}},
  \bibinfo{pages}{033601} (\bibinfo{year}{2002}).


\bibitem[{\citenamefont{G\'{o}ral et~al.}(2004)\citenamefont{G\'{o}ral,
 , and Julienne}}]{Goral2004}
 \bibinfo{author}{\bibfnamefont{K.}~\bibnamefont{G\'{o}ral \emph{et al.}}},
\bibinfo{journal}{J. Phys. B}
 \textbf{\bibinfo{volume}{37}}, \bibinfo{pages}{3457}
 (\bibinfo{year}{2004}).

\bibitem[{\citenamefont{K\"ohler et~al.}(2006)\citenamefont{K\"ohler,
  Goral, , and Julienne}}]{Thorsten2006}
  \bibinfo{author}{\bibfnamefont{T.}~\bibnamefont{K\"ohler}},
  \bibinfo{author}{\bibfnamefont{K.}~\bibnamefont{G\'{o}ral}}, ,
  \bibnamefont{and} \bibinfo{author}{\bibfnamefont{P.~S.}
  \bibnamefont{Julienne}}, \bibinfo{journal}{Rev. Mod. Phys.}
  \textbf{\bibinfo{volume}{78}}, \bibinfo{pages}{1311}
  (\bibinfo{year}{2006}).

\bibitem[{\citenamefont{Szyma\'{n}ska
  et~al.}(2005{\natexlab{b}})\citenamefont{Szyma\'{n}ska, G\'{o}ral,
  K\"{o}hler, and Burnett}}]{me_therm}
\bibinfo{author}{\bibfnamefont{M.~H.} \bibnamefont{Szyma\'{n}ska \emph{et al.} }}, 
  \bibinfo{journal}{Phys. Rev. A} \textbf{\bibinfo{volume}{72}},
  \bibinfo{eid}{013610} (\bibinfo{year}{2005}{\natexlab{b}}).

\bibitem[{\citenamefont{K\"ohler et~al.}(2003)\citenamefont{K\"ohler, Gasenzer,
  and Burnett}}]{Thorsten2003}
\bibinfo{author}{\bibfnamefont{T.}~\bibnamefont{K\"ohler}},
  \bibinfo{author}{\bibfnamefont{T.}~\bibnamefont{Gasenzer}}, \bibnamefont{and}
  \bibinfo{author}{\bibfnamefont{K.}~\bibnamefont{Burnett}},
  \bibinfo{journal}{Phys. Rev. A} \textbf{\bibinfo{volume}{67}},
  \bibinfo{pages}{013601} (\bibinfo{year}{2003}).

\bibitem[{\citenamefont{Tarruell et~al.}(2005)\citenamefont{Tarruell,
  Teichmann, Mckeever, Bourdel, Cubizolles, Khaykovich, Zhang, Navon, Chevy,
  and Salomon}}]{Salomon2005}
\bibinfo{author}{\bibfnamefont{L.}~\bibnamefont{Tarruell \emph{et al.} }},
  \bibinfo{journal}{arXiv:cond-mat/0701181}  (\bibinfo{year}{2005}).

\bibitem[{\citenamefont{Yi and Duan}(2006)}]{Duan}
\bibinfo{author}{\bibfnamefont{W.}~\bibnamefont{Yi}} \bibnamefont{and}
  \bibinfo{author}{\bibfnamefont{L.-M.} \bibnamefont{Duan}},
  \bibinfo{journal}{Phys. Rev. A} \textbf{\bibinfo{volume}{73}},
  \bibinfo{eid}{063607} (\bibinfo{year}{2006}).

\bibitem[{\citenamefont{von Stecher and Greene}(2007)}]{Greene}
\bibinfo{author}{\bibfnamefont{J.}~\bibnamefont{von Stecher}} \bibnamefont{and}
  \bibinfo{author}{\bibfnamefont{C.~H.} \bibnamefont{Greene}},
  \bibinfo{journal}{Phys. Rev. Lett.} \textbf{\bibinfo{volume}{99}},
  \bibinfo{eid}{090402} (\bibinfo{year}{2007}).

\bibitem[{\citenamefont{Hanna et~al.}(2006)\citenamefont{Hanna, G\'{o}ral,
  Witkowska, and K\"{o}hler}}]{Hanna2006}
\bibinfo{author}{\bibfnamefont{T.~M.} \bibnamefont{Hanna \emph{et al.}}},
  \bibinfo{journal}{Phys. Rev. A} \textbf{\bibinfo{volume}{74}},
  \bibinfo{eid}{023618} (\bibinfo{year}{2006}).

\bibitem[{\citenamefont{Zwierlein}(2007)}]{Martin}
\bibinfo{author}{\bibfnamefont{M.~W.} \bibnamefont{Zwierlein}},
  \bibinfo{journal}{Private communication}  (\bibinfo{year}{2007}).

\end{thebibliography}

\end{document}